\documentclass [a4paper,fleqn, 12pt]{article}
\usepackage{graphicx}
\usepackage[small]{subfigure,epsfig}

\usepackage {amsmath} \usepackage{amssymb} \usepackage{cite}

\newcommand{\cn}

\begin{document}


\title
{Thermal regimes of high burn-up nuclear fuel rod}

\author
{Nikolai A. Kudryashov, \and Aleksandr V. Khlunov, \and Mikhail A. Chmykhov}

\date{Department of Applied Mathematics, National Research Nuclear University
MEPHI, 31 Kashirskoe Shosse,
115409 Moscow, Russian Federation}




\maketitle

\begin{abstract}
The temperature distribution in the nuclear fuel rods  for high burn-up
is studied. We use the numerical and analytical approaches. It is shown
that the time taken to have the stationary thermal regime of nuclear
fuel rod is less than one minute. We can make the inference that the
behavior of the nuclear fuel rod can be considered as a stationary
task. Exact solutions of the temperature distribution in the fuel rods in
the stationary case are found. Thermal regimes of high burn-up the nuclear
fuel rods are analyzed.

\end{abstract}






\section{Introduction}

Temperature distribution in the fuel rod of nuclear reactor is one of the most important
factors that controls the behavior of fission products in the pellets, the diffusion and
vaporization properties and so on \cite{Amaya_2001}. This dependence has been intensively
studied for prolonged lifetime of existing reactors. There are many papers in which the
authors studied the different aspects of this problem \cite{Yapici_2003, Kim_2006,
Pontedeiro_2008, Fortini_2008, Espinosa_2006, Espinosa_2009}. Yapici et al.\cite{Yapici_2003} investigated
the maximum temperatures in centerline of the fuel rod for different clad outer surface
temperatures, melting points of the fuel materials, temporal heat generation, temperature
distribution in the nuclear fuel rod and temporal variation of the neutronic data during
rejuvenation periods. In \cite{Kim_2006} Kim et al. developed the one-dimensional heat
conduction model to determine the temperatures distribution from the fuel center to cladding
surface in the radial direction. Pontedeiro et al. in \cite{Pontedeiro_2008} presented an
improved lumped---differential formulation for one---dimensional transient heat conduction in
a heat generating cylinder with temperature---dependent thermo---physical properties typical
of high burn-up nuclear fuel rods. Analytical model for the determination of the temperature
distribution in cylindric heater components with characteristics of nuclear fuel rods is
given by Fortini et al. in \cite{Fortini_2008}. Espinosa---Paredas et al. in paper
\cite{Espinosa_2009} explored the applicability of a fuel rod mathematical model based on
Non---Fourier transient heat conduction as constitutive law for the Light Water Reactors
transient analysis.

The fuel behavior is affected by the temperature distribution in the
fuel that is related to change in the fuel microstructure with
irradiation \cite{Lee_2001}. One significant change in the fuel
microstructure is the formation of a porous rim in the periphery of
the high burn-up fuel \cite{Walker_1992, Matzke_1992, Une_1992,
Lassmann_1995, Spino_1996,  Matzke_1997, Kinoshita_1997,
Kinoshita_1998, Walker_1999, Romano_2007}. High burn-up nuclear fuel
rods have been intensively studied last years
\cite{Pontedeiro_2008}. Many papers were published studying of
the rim formation mechanism \cite{Lee_2001, Kashibe_1993,
Nogita_1995}. It was shown that the rim structure is formed through
recrystallization and coarsened pore formation. The subdivided
grains with high angle grain boundaries are the nuclei of
recrystallization,  and then the coarsened pores are formed by the
sweeping out of small pores during grain growth on recrystallization
\cite{Lee_2001, Kashibe_1993, Nogita_1995}. The rim structure can be
described taking into consideration many specific characteristics of
the fuel \cite{Lee_2001} but in this paper we are not going to touch these
interesting questions.

Here  we study the temperature distribution in the
nuclear fuel rod of the reactor taking the influence of the high
burn-up into account. Using the numerical method based on the
difference equation of the heat conductivity we prove that the
stationary behavior of the nuclear reactor is reached for the time
less then one minute. This fact allows us to consider the
temperature distribution using the stationary cases of the heat
conductivity in the nuclear fuel reactor. At solution of the task we
do not use the modern approaches for the nonintegrable differential
equations (see \cite{Kudryashov_2009a, Kudryashov_2009b,
Kudryashov_2009c}). The task is solved by us taking the simple
integration of equations. As a result we find the solution for the
temperature distribution in the nuclear fuel rod in the analytical
form for the stationary behavior of the reactor. Exact solutions
allow us to analyze the temperature distribution in details and to
evaluate the influence of the rim - layer and the zirconium oxide on
the temperature distribution in the nuclear fuel rod.

This paper is organized as follows. In section 2 we present the
statement of the problem and we use the numerical methods to obtain
the energy flow on the surface of cladding. Using the numerical
simulation for solving task in section 3 we observe that the
stationary case of the temperature distribution in the nuclear fuel
rod arises very quickly in comparison with time of the product of
electricity production. So we can consider the stationary
temperature distribution in the nuclear fuel rod. Using this fact in
section 4 we give the stationary statement of the problem and in section 5
we find the exact solution of the temperature distribution
in the nuclear fuel rod taking rim - layer and zirconium oxide into account
for high burn-up fuel. In section 6 we discuss the particularities
of the stationary thermal regimes in nuclear rods. As this takes
place we consider the different cases of nuclear fuel rods: with gap
and without gap; in the case of the burn-up and  without rim -
layer and zirconium and so on.  We can
allow to do it using the analytical solutions.

\section{The statement of the problem for the temperature distribution
in nuclear fuel rod with rim --- layer and zirconium oxide}

\begin{figure}[ht] 
\centerline{\epsfig{file=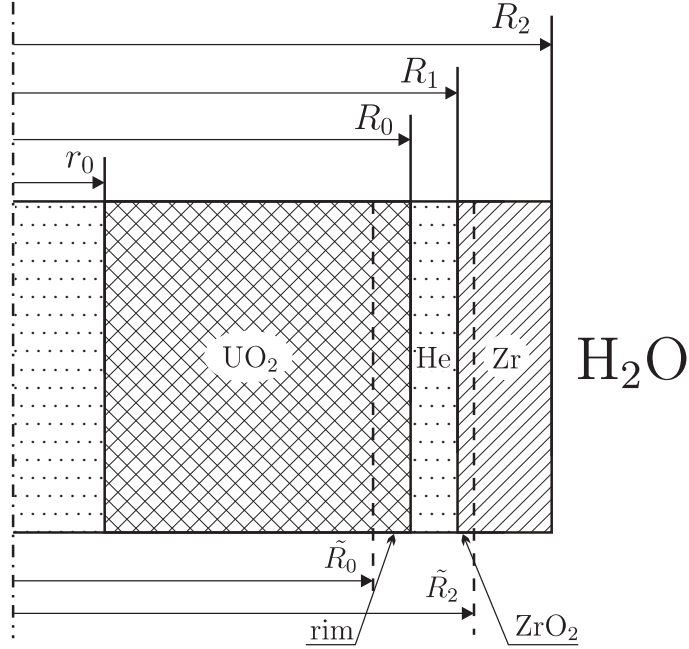,width=100mm}}
\caption{Geometry of the nuclear fuel rod: $r_0$  is central hole,
$R_0$ is radius of the fuel, $\tilde {R}_0$ is the radius of the rim layer, $R_1$ is
external radius of gap,
$\tilde {R}_2$ is radius of $ZrO_2$, $R_2$ is external radius of $Zr$.} \label{fig:_e0a1}
\end{figure}

The geometry of the nuclear fuel rod is given in Fig.
\ref{fig:_e0a1}. We take this geometry into account in this paper to study the temperature
distribution in the nuclear fuel rod.

Let us introduce the following notation: $T_1(r,t)$ is dependence of temperature on $r$
and $t$ in the fuel that is $UO_2$,  $T_2(r,t)$ is the dependence of temperature from radius
and time in the gap, that is $He$,  $T_3(r,t)$ is dependence of temperature on $r$ and $t$ in
the cladding,  $\tilde{T}_1(r,t)$ is temperature in the rim layer, $\tilde{T}_3(r,t)$ is the
temperature in the zirconium oxide, $r_0$ is the radius of the central hole,  $R_0$ is the radius of the nuclear
fuel, $R_1$ is the external radius of the gap in the fuel rod, $R_2$ is the radius of the cladding, $\tilde{R}_0$ is
the internal radius of the rim -
 layer, $\tilde R_2$ is the external radius of the zirconium oxide.

 Nonlinear differential equation for the description of the temperature in the nuclear fuel
 rod can be written as
\begin{equation}\label{eq0a}
{C_p}(r,T)\,{\rho}(r,T)\,\frac{\partial {T}}{\partial t} = \frac{1}{r}
\frac{\partial}{\partial r} \left( \lambda(r,T)\,r \frac{\partial
{T}}{\partial r} \right) + q(r), \qquad {r}_0 \leq r \leq R_2,
\end{equation}
where $C_p(r,T)$ is the specific heat capacity, $\rho(r,T)$ is the mass density,
$\lambda(r,T)$ is the thermal conductivity, $q(t)$ is the  uniform volumetric heat
generation rate which is independent of the temperature.

We take the initial condition for the temperature in the form
\begin{equation}\label{eq1a}
T(r,t=0) = \varphi_{{0}}.
\end{equation}
We used different initial conditions for the numerical simulation taking into account the
initial behavior of nuclear reactor.

The boundary values for the numerical simulation we took at $r=r_0$ in the form
\begin{equation}\label{eq2a}
\left.\frac{\partial T}{\partial r}\right|_{r=r_0} = 0,
\end{equation}
and at $r=R_2$
\begin{equation}\label{eq3a}
T(r=R_2,t) = T_{{w}},
\end{equation}
where $T_w$ is the  temperature of the wall.

Heat capacity $(C_p)$ of oxide fuels is very important parameter for
evaluation of fuel temperature at normal, transient and accidental
conditions of light water reactor \cite{Amaya_2001}. Heat capacities
of undoped and impurity-doped $UO_2$  pellets have been measured by
many researchers. The values for densities and heat capacities of
nuclear fuel, rim layer, gas in gap and in cladding are different
from each other. We take formulae for these values as sectionally -
continuous function
$$
\rho(r,T)\, C_p(r,T) =\begin{cases}\rho_1(T)\,C_{p,1} (T), &\text{if
$r_0\leq r<\tilde{R_0}$},\\
\tilde{\rho_1}(T)\,\tilde{C_{p,1}}(T), &\text{if
$\tilde{R_0}\leq r<{R_0}$},\\
\rho_2\,C_{p,2}, \, &\text{if
${R_0}\leq r<{R_1}$},\\
 \tilde{\rho_3}\, \tilde{C_{p,3}}, &\text{if
$R_1\leq r<\tilde{R_2}$},\\
 \rho_3\, C_{p,3}, &\text{if
$\tilde{R_2}\leq r<R_2$}.\\\end{cases}
$$
We have taken physical parameters and dependencies of the densities
and the specific heat capacities on temperature for the fuel, the
gap and the cladding from the papers \cite{Amaya_2001,
Espinosa_2006, Espinosa_2009} .

Heat conductivities for the fuel and for the rim layer could be expressed  by a classical
phonon transport model \cite{Minato_2001, Ronchi_2004, Mustafa_2004, Morimoto_2008}
\begin{equation}\label{eq_0b}
\lambda_1(T)=\frac{1}{A+c\,bu+B\,T},
\end{equation}
where $A$, $B$ and $c$ are constants and $bu$ is the local burn-up
in the nuclear fuel and in the rim - layer.

For numerical simulation of the temperature distribution in the
nuclear fuel and in the rim - layer We assume that the coefficient
of the heat conductivity is expressed by the sectionally -
continuous function
$$
\lambda(T,r) =\begin{cases}\frac{1}{A+B\,T}, &\text{if
$r_0 \leq r< \tilde {R}_0$}\\
\frac{1}{A_1+B_1\,T}, &\text{if $\tilde {R }_0\leq r< R_0$}\\
\lambda_1, &\text{if $R_0\leq r< R_1$}\\
\tilde{\lambda}_2, &\text{if $R_1\leq r< \tilde {R}_2
$}\\
\lambda_2, &\text{if $\tilde {R}_2\leq r< R_2$,}\end{cases}
$$
where $A$, $B$, $A_1$, $B_1$, $\lambda_1$, $\tilde{\lambda_1}$, $\tilde{\lambda_2}$,
$\lambda_2$ are constants. For numerical simulation of task we have taken values of
these parameters from papers \cite{Minato_2001, Ronchi_2004, Mustafa_2004, Morimoto_2008}.

The volumetric heat generation rate  for the numerical simulation can be written as
$$
q(r) =\begin{cases}q_0, &\text{if
$r_0 \leq r<\tilde R_0$}\\
q_1, &\text{if $\tilde R_0 \leq r< R_0 $}\\
0, &\text{if $ R_0 \leq r< R_2$,}\end{cases}
$$
where $q_0$ and $q_1$ are constants.
Task \eqref{eq0a} - \eqref{eq3a} is solved by us in the next section
using the numerical methods.

\section{The numerical simulation of the temperature distribution in the nuclear fuel rod}

Consider the equation
\begin{equation}\label{eq0aa}
{C_p}(r,T)\,{\rho}(r,T)\,\frac{\partial {T}}{\partial t} = \frac{1}{r}
\frac{\partial}{\partial r} \left( \lambda(r,T)\,r \frac{\partial
{T}}{\partial r} \right) + q(r), \qquad {r}_0 \leq r \leq R_2,
\end{equation}

Multiplying \eqref{eq0aa} on $r$ and integrating with respect to $r$
at constant $q$ and $q_1$, we have
\begin{equation}\label{eq0ab}
\frac{\partial {E}}{\partial t} = 2\,\pi\,\int_{r_0}^{R_2} r
\lambda(r,T) \, \frac{\partial {T}}{\partial r} dr  +\pi q
\,(\tilde{R}_0^{2}-r_{0}^{2})+\pi\,q_1\,(R_{0}^{2}-\tilde R_0^{2}),
\end{equation}
where
$$E=2\pi\,\int_{r_0}^{R_2}{C_p}(r,T)\,{\rho}(r,T)\,T(r,t)\,dr$$

Let us denote the flow of energy via the surface unit as $W$, then Eq.\eqref{eq0aa}
can be written as
\begin{equation}\label{eq0c}
\frac{\partial {E}}{\partial t} = -2\,\pi\,R_2^2\,W  +\pi q \,(\tilde{R}_0^{2}-
r_{0}^{2})+\pi\,q_1\,(R_{0}^{2}-\tilde R_0^{2}),
\end{equation}

We obtain that the flow of energy for stationary behavior of the
nuclear fuel rod takes the form
\begin{equation}\label{eq0ad}
W = \frac{1}{2\,R_2^2}\left(q \,(\tilde{R}_0^{2}-r_{0}^{2})+q_1\,(R_{0}^{2}-
\tilde R_0^{2})\right),
\end{equation}

To solve the task \eqref{eq0a} - \eqref{eq3a} we use the
numerical simulation. With this aim we introduce the grid on
coordinate $r$ and time $t$. We take points on $r$ and $t$ using the
formulae
\begin{equation}\label{eq14}
r_j=j\,h,\qquad (j=0,...J); \qquad t^n=\tau\,n,\qquad (n=0,... N),
\end{equation}
where $h$ and $\tau$ are steps on coordinate and time, $J$
corresponds to the radius $R_2$, $N$ corresponds to the stationary
state of the nuclear fuel rod.

Let us introduce the grid functions at $t=t^n$ and $r=r_j$. In
this case we have
\begin{equation}\begin{gathered}\label{eq14a}
T(r_j,t^n)\simeq T_j^n , \qquad q_j^n\simeq q(r_j,t^{n}), \\
\\
\lambda_j^n\simeq \lambda(T_j^n,\,r_j), \qquad
C_p(T_{j}^{\,n},\,r_j)\simeq C_{p,j}^{\,n}, \qquad \rho(T_{j}^{n}, \,r_j)\simeq \rho_j^{n}.
\end {gathered}\end{equation}

We change the differential operators in Eq.\eqref{eq0a} on difference
operators using the formulae
\begin{equation}\label{eq14aa} \left.\frac{\partial T}{\partial
t}\right|_{r=hj;\,t=\tau n+\frac{\tau}{2}}\simeq\frac{T_j^{n+1}-T_j^{n}}{\tau}, \quad \left.\frac{\partial T}{\partial
r}\right|_{r=hj+\frac{h}{2};\,t=\tau(n+1)}\simeq
\frac{T_{j+1}^{n+1}-T_{j}^{n+1}}{h},
\end{equation}\

Taking the grid functions \eqref{eq14a} and the approximation of
differential relations \eqref{eq14aa} we have the following
difference equation which is approximately
equivalent to \eqref{eq0a}
\begin{equation}\begin{gathered}
\label{eq14c}
\frac{T_j^{n+1}-T_j^{n}}{\tau}=\Lambda_{j+\frac12}^{n}\,\sqrt{\frac{r_{j+1}}{{r_j}}}\,
\left(\frac{T_{j+1}^{n+1}-T_{j}^{n+1}}{h^2}\right)-\\
\\
-\Lambda_{j-\frac12}^{n}\,\sqrt{\frac{r_{j-1}}{{r_j}}}\,
\left(\frac{T_{j}^{n+1}-T_{j-1}^{n+1}}{h^2}\right)+f_{j}^{n}, \qquad f_{j}^{n}=
\frac{q_{j}^{n}}{C_{p,j}^{n}\,\rho_{j}^{n}}, \\
\\
(j=1,...J-1),
\end{gathered}\end{equation}
where we use notation
\begin{equation} \begin{gathered}
\label{eq14d}\Lambda_{j+\frac12}^{n}=\frac{\lambda_j^{n}+\lambda_{j+1}^{n}}{2\, C_{p,j}^{\,n}\,\rho_{j}^{n}},
\qquad\Lambda_{j-\frac12}^{n}=\frac{\lambda_j^{n}+\lambda_{j-1}^{n}}{2\, C_{p,j}^{\,n}\,\rho_{j}^{n}},\qquad
(j=1,...J-1).
\end{gathered}\end{equation}

Difference equation \eqref{eq14c} can be written as
\begin{equation}
\label{eq15a}A_{j}^{n}\,T_{j+1}^{n+1}-D_{j}^{n}\,T_{j}^{n+1}+B_{j}^{n}\,T_{j-1}^{n+1}=F_{j}^{n},
\end{equation}
where
\begin{equation}
\label{eq15b}A_{j}^{n}=\frac{\tau\,}{h^2}\sqrt{\frac{r_{j+1}}{r_j}}\,
\left(\frac{\lambda_j^{n}+\lambda_{j+1}^{n}}{2\,C_{p,j}^{n}\,\rho_{j}^{n}}\right),
\end{equation}

\begin{equation}
\label{eq15c}D_{j}^{n}=1+\frac{\tau\,}{h^2}\sqrt{\frac{r_{j+1}}{r_j}}\left(\frac{\lambda_j^{n}+
\lambda_{j+1}^{n}}{2\,C_{p,j}^{n}\,\rho_{j}^{n}}\right)+
\frac{\tau\,}{h^2}\sqrt{\frac{r_{j-1}}{r_j}}\left(\frac{\lambda_j^{n}+
\lambda_{j-1}^{n}}{2\,C_{p,j}^{n}\,\rho_{j}^{n}}\right),
\end{equation}

\begin{equation}
\label{eq15d}B_{j}^{n}\,=\frac{\tau\,}{h^2}\sqrt{\frac{r_{j-1}}{r_j}}
\left(\frac{\lambda_j^{n}+\lambda_{j-1}^{n}}{2\,C_{p,j}^{n}\,\rho_{j}^{n}}\right),
\end{equation}

\begin{equation}
\label{eq15e}F_{j}^{n}=-T_{j}^{n}-\tau\,f_{j}^{n}.
\end{equation}

Difference equations \eqref{eq15a} are the system of the algebraic
equations which corresponds to the differential equation
\eqref{eq0a}. This system of equations allows us to find
the temperature distribution taking the boundary values
and initial conditions.

We assume that the temperature distribution on time  $t^n = n \tau$ is known.
The task is to solve the system of the algebraic equations and to find the
temperature distribution at the moment $t^{n+1} = (n+1) \tau$.

From the initial condition we know that $T(r, t=0) = \varphi_0$ and
consequently we know the temperature distribution on the radius at
$n=0$
\begin{equation}\label{al01}
T^0_j = \varphi_0
\end{equation}

Taking into consideration the boundary conditions \eqref{eq2a} and
\eqref{eq3a}, we have
\begin{equation}\label{al02}
T^n_1 = T^n_0, \qquad T^n_J = T_l, \qquad \left( n = 0, \ldots, N \right)
\end{equation}

Using the temperature distribution on the first layer on time
\eqref{al01} and conditions \eqref{al02}, we solve the system of the
algebraic equations  \eqref{eq15a} and find the temperature on the
first layer at $t^1 = \tau$. Then taking into consideration values
$T^1_j$, we obtain values of temperature $T^2_j$ at  $t^2 = 2 \tau$
and so on.

The system of algebraic equations \eqref{eq15a} can be solved by the sweep method.

\begin{figure}[ht] 
\centerline{\epsfig{file=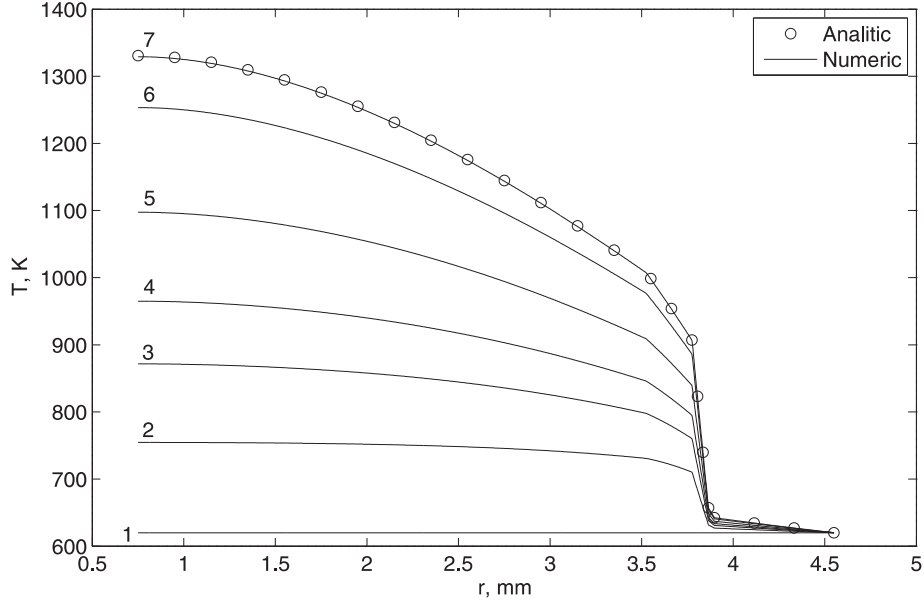,width=140mm}}
\caption{Evolution of temperature in nuclear fuel rod for high burn-up for
$q_L=200$ W/cm. 1:  $t=0$ s; 2:  $t=1$ s; 3:  $t=2$ s; 4:  $t=3$ s;
5:  $t=5$ s; 6:  $t=10$ s; 7:  $t=30$ s;}
\label{fig:f1}
\end{figure}

On figure \ref{fig:f1} we can see the evolution on temperature in
nuclear fuel rod in the case $q_L=200$ W/cm.  Here and later  $q_L\,=q_0\,\pi\,({\tilde{R}}_0^2-\,r_0^2)+q_1\,\pi\,(R_0^2-\tilde{R_0^2})$, where $q_1= 2\, q_0$. We have used the
initial condition for the temperature assuming that this temperature
is equal to temperature of the wall
\begin{equation}\label{al03}
T(r,t=0) = T_w
\end{equation}
Other parameters of the mathematical model are used from the table 1.

From figure \ref{fig:f1} we can see that the stationary state
of the nuclear fuel rod with burn-up is reached during 30 secund of time.

\begin{figure}[ht] 
\centerline{\epsfig{file=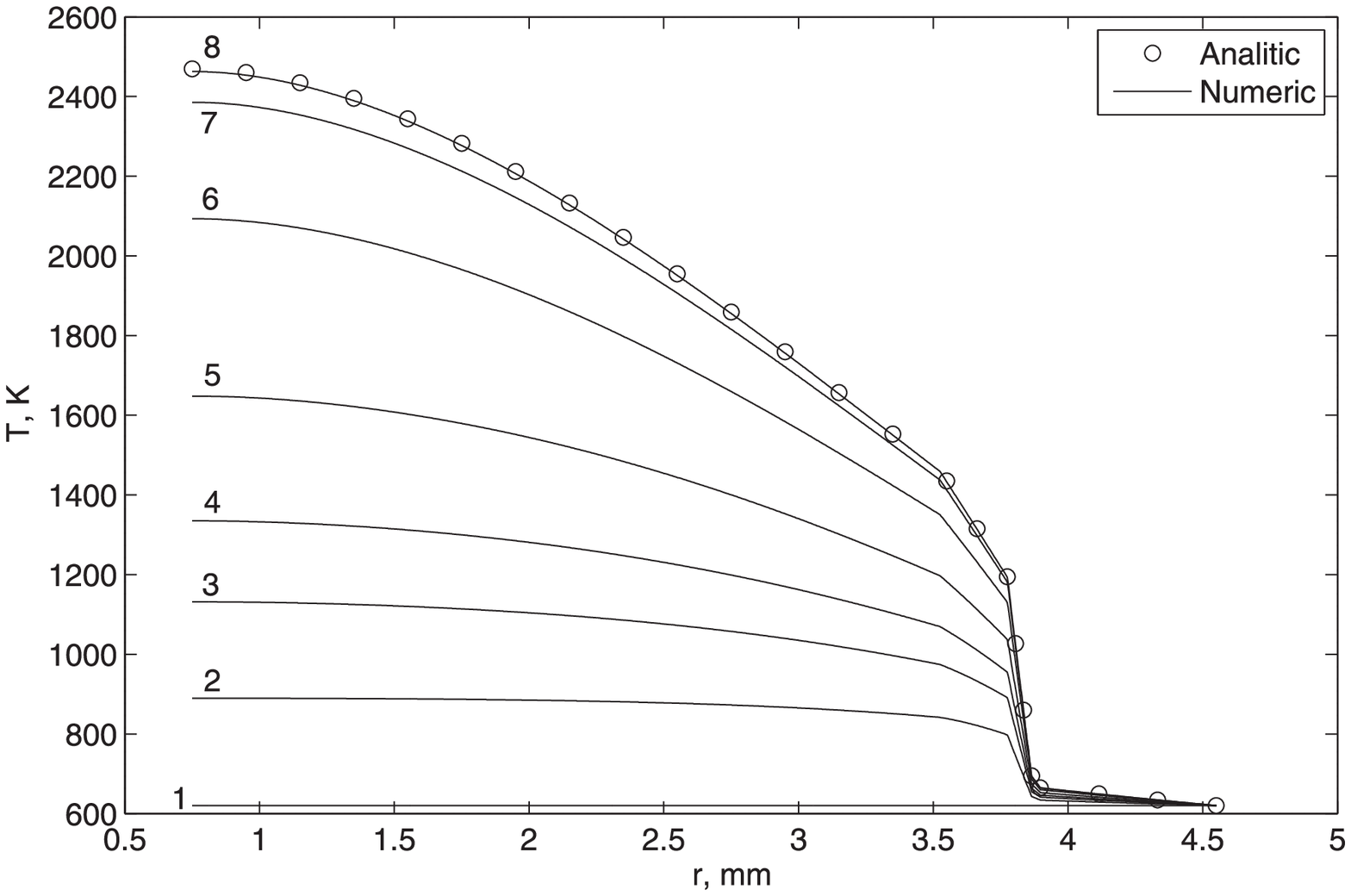,width=140mm}}
\caption{Evolution of temperature in the nuclear fuel rod for high burn-up
at $q_L=400$ W/cm. 1: $t=0$ s; 2:  $t=1$ s; 3:  $t=2$ s; 4: $t=3$ s;
5:  $t=5$ s; 6:  $t=10$ s; 7:  $t=20$ s; 8:  $t=40$ s;}
\label{fig:f2}
\end{figure}

Figure \ref{fig:f2} demonstrates the output the temperature
to the stationary regime of the nuclear fuel rod at the different values
of the volumetrical source rate. We can observe that the time of the output
on the stationary state is not more then 30 seconds. Therefore the basic
temperature regimes of the nuclear fuel rods are the stationary
regimes and we can consider the temperature distribution
in the nuclear fuel rod as the stationary task.

\begin{figure}[ht] 
\centerline{\epsfig{file=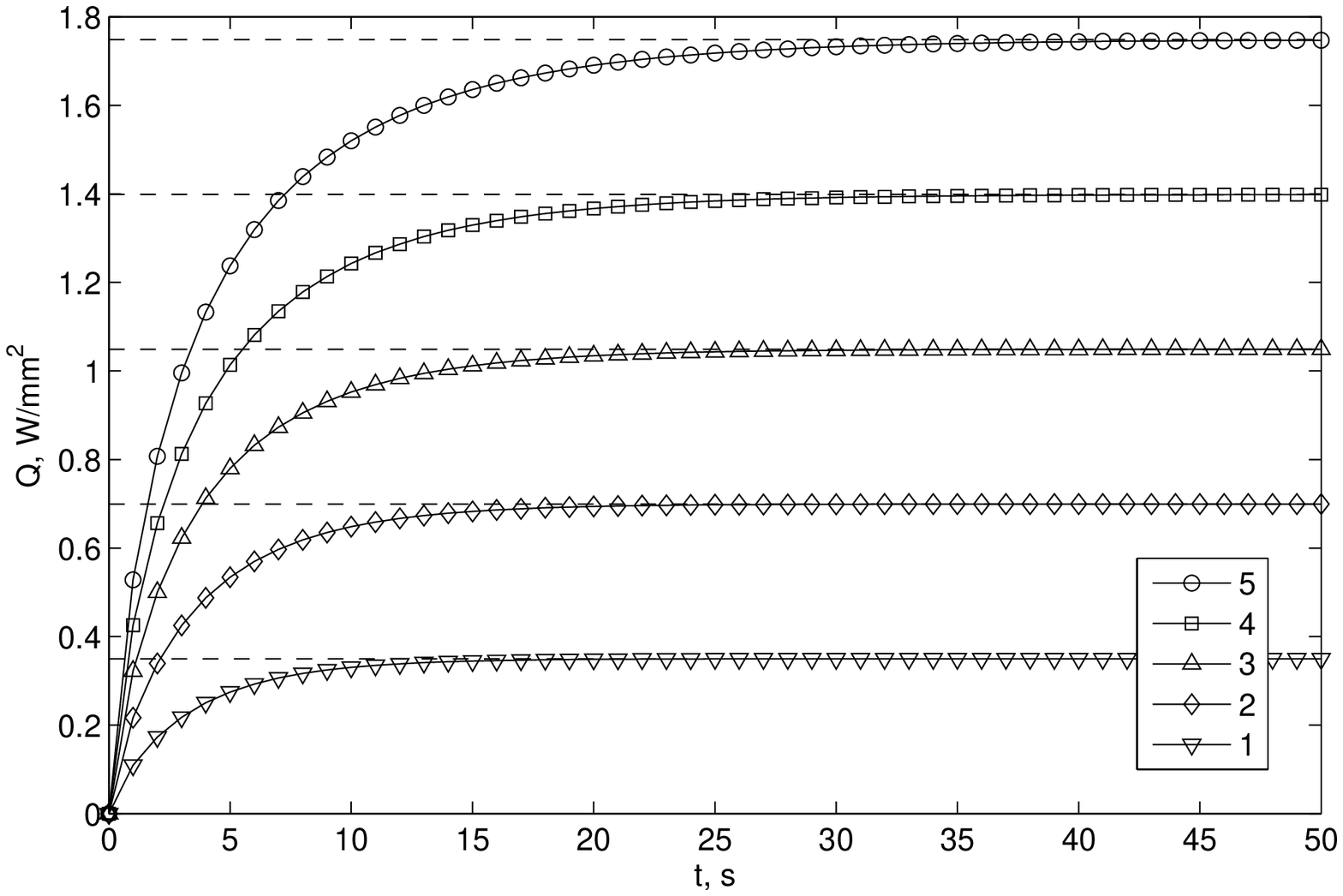,width=140mm}}
\caption{Dependence of the heat flow on time via surface at $r=R_2$
for sources  1:  $q_L=100$ W/cm; 2:  $q_L=200$ W/cm; 3:
$q_L=300$ W/cm; 4:  $q_L=400$ W/cm; 5:  $q_L=500$ W/cm; ($bu=120$)}
\label{fig:f3}
\end{figure}

The numerical simulation of dependence for the heat flow on time at
$r=R_2$  is given on figure \ref{fig:f3}. From this figure we can
see that the time from the initial state to the stationary behavior
is  40 seconds. This time depends on the value of source and
the time grows with increasing source. However this time is not
more 45 seconds for available sources in nuclear fuel rod.

\section{The statement of the stationary problem for the  temperature distribution in
nuclear fuel rods with the rim - layer and the zirconium oxide}

Using the numerical simulation we have obtained in the previous section that the
basic behavior of the nuclear fuel rood is stationary. Consequently we can consider the temperature distribution in
the nuclear fuel rod taking the above mentioned problem when $\frac{\partial T}{\partial t}=0 $. In this case we
have the following statement of the problem for the temperature distribution
\begin{equation}\label{stat01}
\frac{1}{r} \frac{d}{d r} \left( \frac{r}{A+B T_1} \frac{d T_1}{d r} \right) + q = 0,
\qquad r_0 < r < \tilde{R}_0
\end{equation}

\begin{equation}\label{stat01a}
\frac{1}{r} \frac{d}{d r} \left( \frac{r}{A_1+B_1 \tilde{T}_1} \frac{d \tilde{T}_1}{d r}
\right) + q_1 = 0,
\qquad \tilde{R}_0 < r < R_0
\end{equation}

\begin{equation}\label{stat02}
\frac{1}{r} \frac{d}{d r} \left(r \lambda_1 \frac{d T_2}{d r} \right) = 0,
\qquad R_0 < r < R_1
\end{equation}

\begin{equation}\label{stat03}
\frac{1}{r} \frac{d}{d r} \left(r \tilde{\lambda}_2 \frac{d \tilde{T}_3}{d r} \right) = 0,
\qquad R_1 < r < \tilde{R}_2
\end{equation}

\begin{equation}\label{stat03a}
\frac{1}{r} \frac{d}{d r} \left(r \lambda_2 \frac{d T_3}{d r} \right) = 0,
\qquad \tilde{R}_2 < r < R_2
\end{equation}
where $T_1(r)$ is the dependence of the temperature on radius in the
fuel, $\tilde{T}_1(r)$ is the dependence of the temperature on radius in the
rim layer, $T_2(r)$ is the dependence of the temperature on radius in the
gap, $\tilde{T}_3(r)$ is the dependence of the temperature on radius in the
zirconium oxide, $T_3(r)$ is the dependence of the temperature on radius in the
cladding, $q$ and $q_1$ are volumetric sources in the fuel and in
the rim - layer.

To obtain the solution of the system of equations \eqref{stat01} - \eqref{stat03a} we have to
add the boundary values for this system of equations. In this case we use the boundary
conditions at $r=r_0$, $r=\tilde{R}_0$, $r=R_0$, $r=R_1$ $r=\tilde{R}_2$ and at $r=R_2$.

The boundary value for the temperature $T_1(r)$ at $r=r_0$ takes the form
\begin{equation}\label{stat04}
\left. \frac{d T_1}{d r} \right|_{r=r_0} = 0.
\end{equation}
We have conditions at $r=\tilde{R}_0$ in the form
\begin{equation}\begin{gathered}\label{stat05}
\frac{1}{A+B T_1} \left.\frac{d T_1}{d r}\right|_{r=\tilde{R}_0} =\frac{1}{A_1+B_1
\tilde{T}_1}
\left.\frac{d \tilde{T}_1}{d r}\right|_{r=\tilde{R}_0}, \\
\\
\quad T_1(r=\tilde{R}_0) = \tilde{T}_1(r=\tilde{R}_0).
\end{gathered}\end{equation}

In the case $r=R_0$, we use the following conditions
\begin{equation}\label{stat05a}
\frac{1}{A_1+B_1 \tilde{T}_1} \left.\frac{d \tilde{T}_1}{d r}\right|_{r=R_0} =
\lambda_1 \left.\frac{d T_2}{d r}\right|_{r=R_0}, \quad \tilde{T}_1(r=R_0) = T_2(r=R_0).
\end{equation}
We take the conditions at $r=R_1$
\begin{equation}\label{stat06}
\lambda_1 \left.\frac{d T_2}{d r}\right|_{r=R_1} =
\tilde{\lambda}_2 \left.\frac{d \tilde{T}_3}{d r}\right|_{r=R_1}, \qquad T_2(r=R_1) =
\tilde{T}_3(r=R_1).
\end{equation}
Assuming the flow at $r=\tilde{R}_2$ we obtain the conditions
\begin{equation}\label{stat06a}
\tilde{\lambda}_2 \left.\frac{d \tilde{T}_3}{d r}\right|_{r=\tilde{R}_2}=\lambda_2 \left.
\frac{d T_3}{d r}\right|_{r=\tilde{R}_2}, \qquad \tilde{T}_3(r=\tilde{R}_2) =
{T}_3(r=\tilde{R}_2).
\end{equation}

We believe that  temperature on the surface of the cladding at $r=R_2$ is constant and this
temperature is equals to the  temperature on the wall $T_w$ \cite{Espinosa_2006}
\begin{equation}\label{stat04b}
 T_3(r=R_2)= T_w.
\end{equation}

System of equations \eqref{stat01} - \eqref{stat03a} and conditions
\eqref{stat04} - \eqref{stat04b} describe the stationary temperature
in the nuclear fuel rod.

\section{Exact solutions for the temperature distribution in the
stationary nuclear fuel rod with rim - layer and film of zirconium oxide}

Solutions of the system of equations \eqref{stat01} ---
\eqref{stat04b} can be obtained in the analytical form. Integrating
equation \eqref{stat01} with respect to $r$, we have
\begin{equation}\label{S03b}
\frac{r}{A+B\,T_1}\,\frac{dT_1}{dr}+\frac{q\,r^2}{2}=C_1,
\end{equation}
where $C_1$ is arbitrary constant.

Eq.\eqref{S03b} can be written as
\begin{equation}\label{S04c}
\,\frac{d \ln{(A+B\,T_1)}}{dr}+\frac{B\,q\,r}{2}=\frac{B\,C_1}{r}.
\end{equation}

Integrating Eq.\eqref{S04c} with respect to $r$ again we obtain
\begin{equation}\label{S05d}
\ln{(A+B\,T_1)}= \ln(C_2)+{B\,C_1}\ln{(r)}-\frac{B\,q\,r^2}{4},
\end{equation}
where $C_2$ is arbitrary constant as well. From Eq.\eqref{S05d} we have
the solution of Eq.\eqref{stat01}
\begin{equation}\label{R01}
T_1(r)= \frac{C_2}{B}\,r^{B\,C_1}\,e^{\frac{-B\,q\,r^2}{4}}-\frac{A}{B}.
\end{equation}
By analogue we find the solution of Eq. \eqref{stat01a} for
dependence temperature on radius in rim layer
\begin{equation}\label{R02}
\tilde{T}_1(r)= \frac{\tilde{C}_2}{B_1}\,r^{B_1\,\tilde{C}_1}\,e^{\frac{-
B_1\,q_1\,r^2}{4}}-
\frac{A_1}{B_1},
\end{equation}
where $\tilde{C_1}$ and $\tilde{C_2} $ are constants of integration as well.

The general solutions of Eqs. \eqref{stat02} - \eqref{stat03a} are
determined by formulae
\begin{equation}\label{R03}
T_2(r) = \frac{C_3}{\lambda_1} \ln(r) + C_4,
\end{equation}
\begin{equation}\label{R04}
\tilde{T}_3(r) = \frac{\tilde{C}_5}{\tilde{\lambda}_2} \ln(r) + \tilde{C}_6,
\end{equation}
\begin{equation}\label{R05}
T_3(r) = \frac{C_5}{\lambda_2} \ln(r) + C_6,
\end{equation}
where $C_3$,  $C_4$,  $\tilde{C}_5$,  $\tilde{C}_6$,  $C_5, $  and  $C_6$ are arbitrary
constants as well.

In order to find the temperature distribution in the nuclear fuel
rod we have to obtain values of the arbitrary constants $C_1$,
$C_2$, $\tilde{C}_1$,  $\tilde{C}_2$, $C_3$, $C_4$, $\tilde{C}_5$,
$\tilde{C}_6$,  $C_5, $  and   $C_6$ taking conditions
\eqref{stat04} - \eqref{stat04b} into account.

Taking the boundary value into account
\begin{equation}\label{S08}
\left. \frac{d T_1}{d r} \right|_{r=r_0} = 0,
\end{equation}
we obtain the constant $C_1$ in the form
\begin{equation}\label{S09}
C_1=\frac{q\,r_0^2}{2}.
\end{equation}

Substituting solutions \eqref{R01} - \eqref{R05} into conditions
\eqref{stat04} - \eqref{stat04b}, we have the system of algebraic
equations with respect to constants $C_2$, $\tilde{C}_1$,
$\tilde{C}_2$, $C_3$, $C_4$, $\tilde{C}_5$, $\tilde{C}_6$, $C_5, $ è
$C_6$.  This system of equations takes the form

\begin{equation}\label{stR10}
\frac{C_2}{B} \, \tilde{R}_0^{\frac{q B r_0^2}{2}}\,e^{-\frac{B q \tilde{R}_0^2}{4}}-
\frac{A}{B} = \frac{\tilde{C}_2}
{B_1}\,\tilde{R}_0^{\,B_1 \tilde{C}_1} \, e^{-\frac{B_1q_1\tilde{R}_0^2}{4}}-
\frac{A_1}{B_1},
\end{equation}

\begin{equation}\label{stR11}
\frac{q r_0^2}{2 \tilde{R}_0} - \frac{q \tilde{R}_0}{2} = \frac{\tilde{C}_1}
{\tilde{R}_0} - \frac{q_1 \tilde{R}_0}{2},
\end{equation}

\begin{equation}\label{stR12}
\frac{\tilde{C}_2}{B_1}R_0^{B_1\tilde{C}_1} \, e^{-\frac{B_1q_1R_0^2}{4}}-\frac{A_1}{B_1} =
\frac{C_3}{\lambda_1} \ln R_0 + C_4,
\end{equation}

\begin{equation}\label{stR13}
\frac{\tilde{C}_1}{R_0} - \frac{q_1 R_0}{2} = \frac{C_3}{R_0},
\end{equation}

\begin{equation}\label{stR14_1}
\frac{C_3}{\lambda_1} \ln R_1 + C_4 = \frac{\tilde{C}_5}{\tilde{\lambda}_2}
\ln R_1 + \tilde{C}_6,
\end{equation}

\begin{equation}\label{stR14_2}
 C_3 = \tilde{C}_5,
\end{equation}

\begin{equation}\label{stR14_3}
\frac{\tilde{C}_5}{\tilde{\lambda}_2} \ln \tilde{R}_2 + \tilde{C}_6 =
\frac{C_5}{\lambda_2} \ln \tilde{R}_2 + C_6,
\end{equation}

\begin{equation}\label{stR14_4}
\tilde{C}_5 = C_5,
\end{equation}

\begin{equation}\label{stR14_5}
\frac{C_5}{\lambda_2} \ln R_2 + C_6 = T_w.
\end{equation}

Solving this system of equations with respect to constants $C_2$,
$\tilde{C}_1$, $\tilde{C}_2$, $C_3$, $C_4$, $\tilde{C}_5$,
$\tilde{C}_6$,  $C_5, $ è  $C_6$, we obtain dependencies of them on
parameters of task in the form
\begin{equation}\label{stR20}
\begin{gathered}
C_2 =B \, \left\{T_w +\frac{A_1 }{B_1} + \left(\frac{q \left( r_0^2 - \tilde{R}_0^2
\right)}{2} +
\frac{q_1 \left( \tilde{R}_0^2 - R_0^2 \right)}{2}\right)
\ln {\left[\left( \frac{\tilde{R}_2}{R_1}
\right)^{\frac{1}{\lambda_2}}  \left( \frac{R_1}{R_2}
\right)^{\frac{1}{\tilde{\lambda}_2}} \left( \frac{R_0}{R_1} \right)^{\frac{1}{\lambda_1}} \right]}
\right\}  \cdot \\
\\
\cdot {\left(\frac{\tilde{R}_0}{R_0}\right)^{ \frac{B_1 q \left( r_0^2 - \tilde{R}_0^2
\right)+
B_1q_1  \tilde{R}_0^2 }{2}}} {\tilde{R}_0^{\frac{- q B r_0^2}{2} } }\,{e^{\frac{q_1 B_1
(R_0^2-
\tilde{R}_0^2)+q B \tilde{R}_0^2}{4}}}
+\left( {A} - \frac{A_1 {B}}{B_1} \right) {\tilde{R}_0^{\frac{- q B r_0^2}{2} }}
{e^{\frac{ q B \tilde{R}_0^2}{4}}} ,
\end{gathered}
\end{equation}

\begin{equation}\label{stR16}
\begin{gathered}
\tilde{C}_1 = \frac{q \left( r_0^2 - \tilde{R}_0^2 \right)}{2} + \frac{q_1
\left( \tilde{R}_0^2 \right)}{2} ,
\end{gathered}
\end{equation}

\begin{equation}\label{stR19}
\begin{gathered}
\tilde{C}_2 = \left\{ B_1 \, T_w +A_1 + B_1 C_3 \ln {\left[ \left( \frac{\tilde{R}_2}{R_1}
\right)^{\frac{1}{\lambda_2}}  \left( \frac{R_1}{R_2}
\right)^{\frac{1}{\tilde{\lambda}_2}} \left( \frac{R_0}{R_1} \right)^{\frac{1}{\lambda_1}} \right]} \right\}
{R_0^{- B_1
\tilde{C}_1}} {e^{\frac{B_1q_1  R_0^2}{4}}},
\end{gathered}
\end{equation}

\begin{equation}\label{stR15-1}
\begin{gathered}
C_3 = \frac{q \left( r_0^2 - \tilde{R}_0^2 \right)}{2} + \frac{q_1 \left( \tilde{R}_0^2 -
R_0^2 \right)}{2},
\end{gathered}
\end{equation}

\begin{equation}\label{stR18}
\begin{gathered}
C_4 =  T_w - \left(\frac{q \left( r_0^2 - \tilde{R}_0^2 \right)}{2} + \frac{q_1
\left( \tilde{R}_0^2 -
R_0^2 \right)}{2}\right)\ln {\left[\left( \frac{{R}_2}{\tilde{R}_2} \right)^{\frac{1}
{{\lambda}_2}}
\left( \frac{\tilde{R}_2}{R_1} \right)^{\frac{1}{\tilde{\lambda}_2}}
R_1^{\frac{1}{\lambda_1}}
\right]} ,
\end{gathered}
\end{equation}

\begin{equation}\label{stR15-2}
\begin{gathered}
\tilde{C}_5 = \frac{q \left( r_0^2 - \tilde{R}_0^2 \right)}{2} + \frac{q_1
\left( \tilde{R}_0^2 - R_0^2 \right)}{2},
\end{gathered}
\end{equation}

\begin{equation}\label{stR17}
\begin{gathered}
\tilde{C}_6 = T_w - \left(\frac{q \left( r_0^2 - \tilde{R}_0^2 \right)}{2} + \frac{q_1
\left( \tilde{R}_0^2 - R_0^2 \right)}{2}\right)\, \ln {\left[\left( \frac{{R}_2}
{\tilde R_2} \right)^{\frac{1}{{\lambda}_2}}\left(\tilde R_2^{\frac{1}
{\tilde\lambda_2}} \right)\right]} .
\end{gathered}
\end{equation}

\begin{equation}\label{stR15-3}
\begin{gathered}
C_5 = \frac{q \left( r_0^2 - \tilde{R}_0^2 \right)}{2} + \frac{q_1
\left( \tilde{R}_0^2 - R_0^2 \right)}{2},
\end{gathered}
\end{equation}

\begin{equation}\label{stR17-2}
\begin{gathered}
C_6 = T_w - \left(\frac{q \left( r_0^2 - \tilde{R}_0^2 \right)}{2} + \frac{q_1
\left( \tilde{R}_0^2 -
R_0^2 \right)}{2}\right)\, \ln {\left( R_2^{\frac{1}{\lambda_2}} \right)} .
\end{gathered}
\end{equation}

Substituting these constants into the general solutions of the
system of equations we have formulae to determine the temperature
distribution in the nuclear fuel rod. The temperature distribution
in the fuel without rim - layer takes the form
\begin{equation}\label{stRZr01}
\begin{gathered}
T_1(r) = \left\{T_w +\frac{A_1}{B_1} + W_0 \ln {\left[ \left( \frac{r}{R_1}
\right)^{\frac{1}{\lambda_1}} \left( \frac{R_1}{\tilde{R_2}} \right)^{\frac{1}
{\tilde{\lambda_2}}} \left( \frac{\tilde{R_2}}{R_2} \right)^{\frac{1}
{\lambda_2}} \right]} \right\} \cdot \\
\cdot {\left(\frac{\tilde{R}_0}{R_0}\right)^{ B_1 \,W_1}} {\left(\frac{r}
{\tilde{R}_0}\right)^{\frac{q B r_0^2}{2} } } {e^{\frac{q_1 B_1
(R_0^2-\tilde{R}_0^2)+ q B (\tilde{R}_0^2 - r^2)}{4}}} + \\
+ \left( \frac{A}{B} - \frac{A_1}{B_1} \right) {\left(\frac{r}
{\tilde{R}_0}\right)^{\frac {q B r_0^2}{2} }} {e^{ \frac{q B
(\tilde{R}_0^2 - r^2)}{4}}} - \frac{A}{B},
\end{gathered}
\end{equation}
where we use the notation
\begin{equation}\label{stRZr06}
\begin{gathered}
W_0 = \frac{q \left( r_0^2 - \tilde{R}_0^2 \right)}{2} + \frac{q_1
\left( \tilde{R}_0^2 - R_0^2 \right)}{2},
\quad W_1 = \frac{q \left( r_0^2 - \tilde{R}_0^2 \right)}{2} +
\frac{q_1 \tilde{R}_0^2 }{2}.
\end{gathered}
\end{equation}

The temperature distribution in the rim - layer has dependence on
parameters of task and on the radius in the form
\begin{equation}\label{stRZr02}
\begin{gathered}
\tilde{T}_1(r) = \left\{ T_w + \frac{A_1}{B_1} + W_0 \ln {\left[ \left( \frac{r}{R_1}
\right)^{\frac{1}{\lambda_1}} \left( \frac{R_1}{\tilde{R_2}} \right)^{\frac{1}
{\tilde{\lambda_2}}} \left( \frac{\tilde{R_2}}{R_2} \right)^{\frac{1}{\lambda_2}}
\right]} \right\} \cdot \\
\cdot {\left( \frac{r}{R_0}\right)^{B_1 \,W_1}} {e^{\frac{q_1 B_1 (R_0^2 -r^2)}{4}}} -
\frac{A_1}{B_1}.
\end{gathered}
\end{equation}

The temperature distribution in the gap can be written as
\begin{equation}\label{stRZr03}
\begin{gathered}
T_2(r) =  T_w + W_0 \ln {\left[ \left( \frac{r}{R_1} \right)^{\frac{1}{\lambda_1}}
\left( \frac{R_1}{\tilde{R_2}} \right)^{\frac{1}{\tilde{\lambda_2}}}
\left( \frac{\tilde{R_2}}{R_2} \right)^{\frac{1}{\lambda_2}} \right]}.
\end{gathered}
\end{equation}

The temperature in the film of the zirconium oxide  is expressed by
formula
\begin{equation}\label{stRZr04}
\begin{gathered}
\tilde{T}_3(r) =  T_w + W_0\, \ln {\left[ \left( \frac{r}{\tilde{R_2}}
\right)^{\frac{1}{\tilde{\lambda_2}}} \left( \frac{\tilde{R}_2}{R_2} \right)^
{\frac{1}{\lambda_2}} \right]}.
\end{gathered}
\end{equation}

The temperature distribution in the cladding is found by formula
\begin{equation}\label{stRZr05}
\begin{gathered}
T_3(r) =  T_w + W_0\, \ln {\left[ \left( \frac{r}{R_2} \right)^{\frac{1}
{\lambda_2}} \right]}.
\end{gathered}
\end{equation}

The temperature distribution in the nuclear fuel rod is described by
the continuous function with the break of the first derivative
\begin{equation}\label{stat07f}
T(r) =\begin{cases}T_1(r),  & \text{if $r_0\leq r< \tilde {R}_0$}, \\
\tilde T_1(r), & \text{if $\tilde R_0\leq r< R_0$},\\
T_2(r), & \text{if $R_0\leq r< R_1$}, \\
\tilde T_3(r), & \text{if $R_1\leq r<\tilde R_2$}, \\
T_3(r), & \text{if $\tilde R_2\leq r< R_2$}.
\end{cases}
\end{equation}

Using formula \eqref{stat07f} we can find the temperature
distribution in the nuclear fuel rod at given volumetric source rate
in the fuel $q$ and in the rim - layer $q_1$. We can find the
temperature distribution in the case of different thickness of the
film for the zirconium oxide and at different sizes of the rim -
layer. From formulae \eqref{stRZr01} - \eqref{stRZr05} we can obtain
a number of partial dependencies for the temperature distribution in
nuclear fuel rod. We can find the temperature distribution in the
nuclear fuel rod without rim - layer, without the zirconium oxide,
without gap and so on. With this aim we have to take
$\tilde{R_0}=R_0$ (rim - layer is absent), $\tilde{R}_2=R_2$ (the
zirconium oxide is absent), $R_0=R_1$ (gap is absent) and so on.

\section{Results and discussion}

Let us consider the temperature distribution in a nuclear fuel rod taking
the rim - layer and the film of zirconium oxide into account.  To analyze
the temperature distribution we use sizes of nuclear fuel rod from the
table \ref {tab:dem}.
\begin{table}[ht]
    \center
    \caption{Parameters of the nuclear fuel element} \label{tab:dem}
    \begin{tabular}{|c|c|c|c|}     
        \hline 
          $r_0$, mm & $R_0$, mm & $R_1$, mm & $R_2$, mm \\ \hline
          0.75  & 3.775 & 3.865 & 4.550 \\ \hline
    \end{tabular}
\end{table}

Parameters of heat conductivities of the fuel, the helium in the
gap, the film of the zirconium oxide and the cladding of the
zirconium are given in the table \ref {tab:material}. We assume,
that the heat conductivity is 2 times less  in a rim---layer, than
the heat conductivity in the fuel core.
\begin{table}[ht]
    \center
    \caption{Parameters of heat conductivities of fuel element}
    \label{tab:material}
    \begin{tabular}{|c|c|c|c|c|c|}     
        \hline 
         $A$, mm$\cdot$K/W & $B$, mm/W & $\lambda_1$, W/mm$\cdot$K & $\tilde{\lambda_2}$, W/mm$\cdot$K & $\lambda_2$, W/mm$\cdot$K & $T_w$, K \\ \hline
         43.8              & 0.2294    & $0.3\cdot10^{-3}$         & $1.8\cdot10^{-3}$                 & $22.0\cdot10^{-3}$        & 620      \\ \hline
    \end{tabular}
\end{table}

We assume that the temperature of the coolant $T_w$ is equal to $
620 \, \mbox {K} $ ($ \simeq 350 \, ^ {\circ} \mbox {C} $).

\begin{figure}[ht] 
\centerline{\epsfig{file=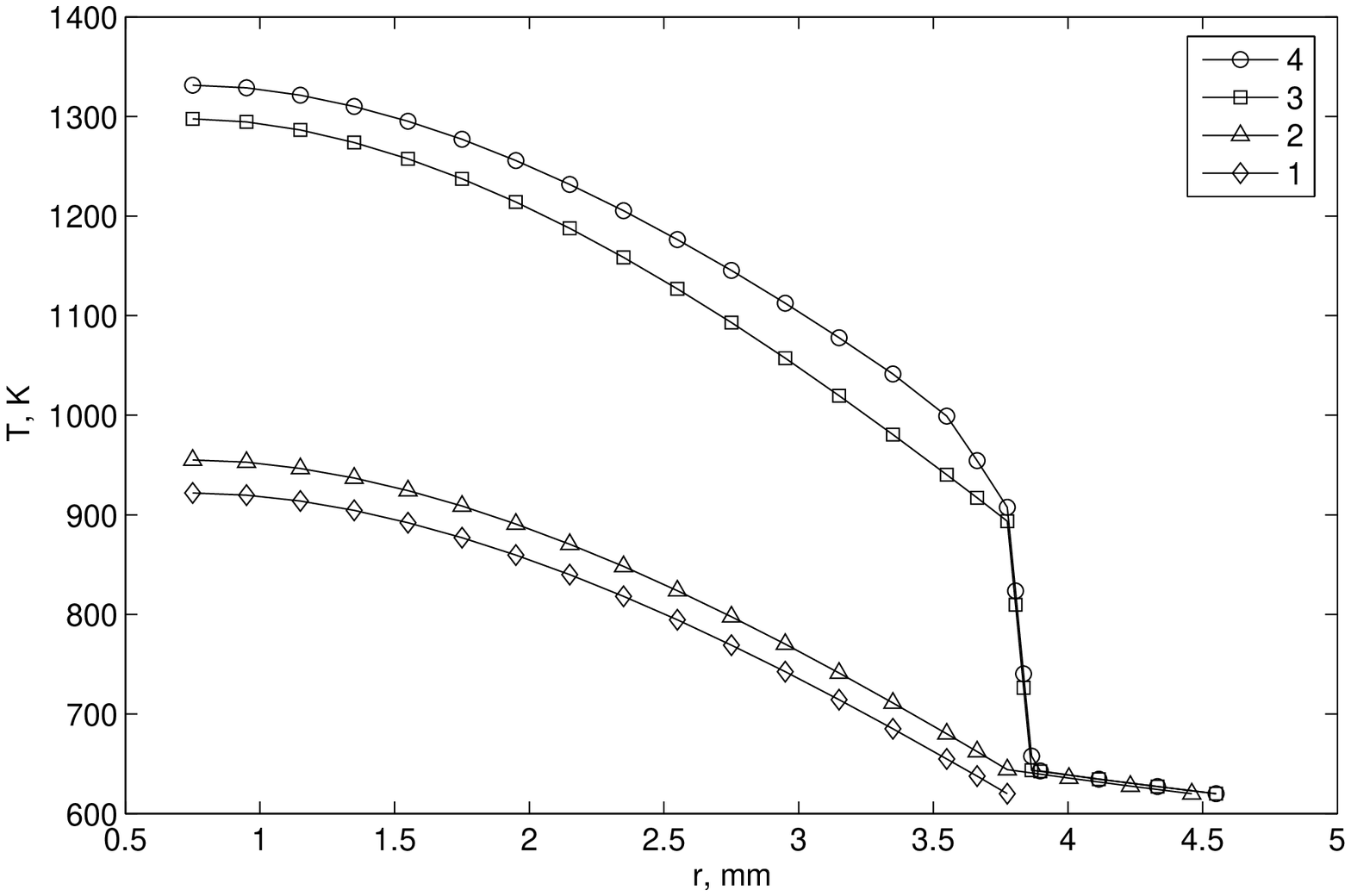,width=140mm}}
\caption{Dependence of temperature on radius at  $q_L=200$~W/cm for cases: 1:  gap
and cladding are out; 2:  gap is absent; 3:  $bu = 0$; 4:  $bu = 120$;}
\label{fig:f4}
\end{figure}

\begin{figure}[ht] 
\centerline{\epsfig{file=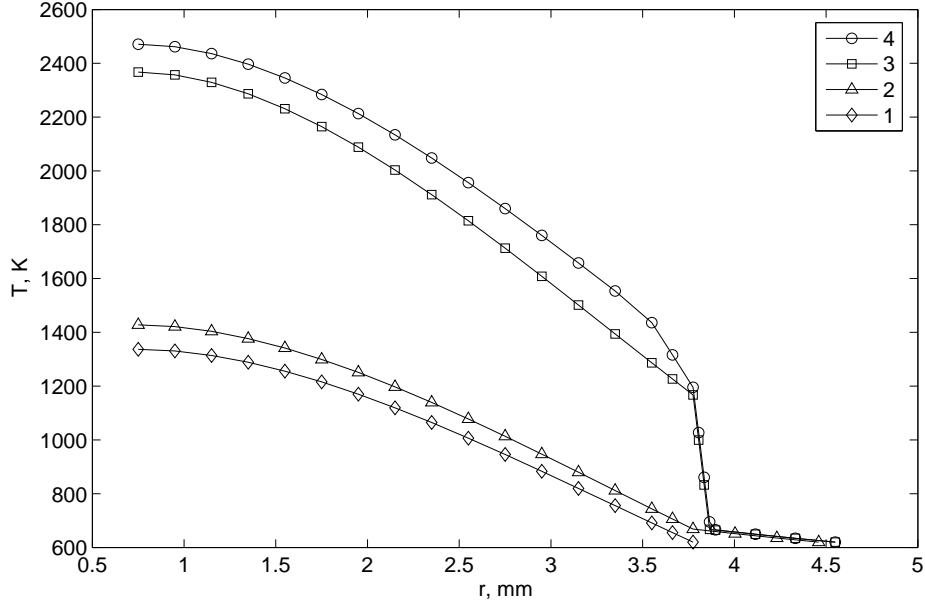,width=140mm}}
\caption{Dependence of temperature on radius at source $q_L=400$~W/cm for cases: 1:
gap and cladding are absent; 2:  gap is absent; 3:  $bu = 0$; 4:  $bu = 120$;}
\label{fig:f5}
\end{figure}

Some thermal regimes of temperature distributions in the nuclear
fuel rod  are given on figure \ref{fig:f4} and on figure
\ref{fig:f5} by the formulae \eqref{stRZr01} --- \eqref{stRZr05} . The
first figure illustrates curves for the value of the source
$q_L=200$ W/cm. The second figure demonstrates the thermal regimes
in the case of the source $q_L=400$ W/cm.

Curve 1 illustrates the temperature distribution when the gap and
the cladding are absent. This case is possible when there is
breakage of the cladding and the coolant has contact with the fuel.
Temperature of the fuel in this case is less then in other cases.
The decreasing of temperature derives from the fact that energy of
the fuel goes to the coolant because the heat conductivity of fuel
is more then the heat conductivity of the helium in the gap.

Curve 2 in figures corresponds to the case with cladding but when the
gap is absent. One can observe that the temperature is higher then in previous
case but this one is less then the temperature in the nuclear fuel
rod for the normal behavior.

Curve 3 illustrates the case of the normal behavior of the nuclear
fuel rod. The temperature distribution is higher of the previous
cases but this one is less then in the case with the formation of
the rim - layer.

Curve 4 on figure \ref{fig:f4} and on figure \ref{fig:f5}
corresponds to the thermal regime in nuclear fuel rod of high
burn-up. We take into account the burn-up equal 120. We can see that
the temperature in this case of the nuclear fuel rod is much more
then other cases. The comparison of the temperature in this case
with the point of the melting fuel shows that there is dangerous
behavior of the nuclear reactor because it is possible the melting
of the fuel at formation of the rim - layer. This fact illustrates
that in the case of increasing of the rim - layer we have to
decrease the value of the source.

At numerical calculations we suppose that the film of the zirconium
oxide is formed at the expense of the gap and the cladding at
various values of burn-up of the fuel. As this takes place the
thickness of the gap is decreased. We assume that the film of
the zirconium oxide takes 90 \% of the cladding and 10 \% of the
volume in the gap.

\begin{table}[ht]
    \center
    \caption{Parameters at burn-up of fuel} \label{tab:t04B}
    \begin{tabular}{|l|c|c|c|c|c|}     
        \hline 
          Burn-up, GWt $\cdot$ twenty four hours/t$\cdot$U & 0 & 60 & 80    & 100   & 120  \\ \hline
          Thickness of a layer ZrO$_2$, mm     & 0 & 0.0168 & 0.0224 & 0.028 & 0.0336 \\ \hline
          Thickness of a rim --- layer, mm     & 0 & 0.1    & 0.15   & 0.2   & 0.25   \\ \hline
    \end{tabular}
\end{table}

\begin{figure}[ht] 
\centerline{\epsfig{file=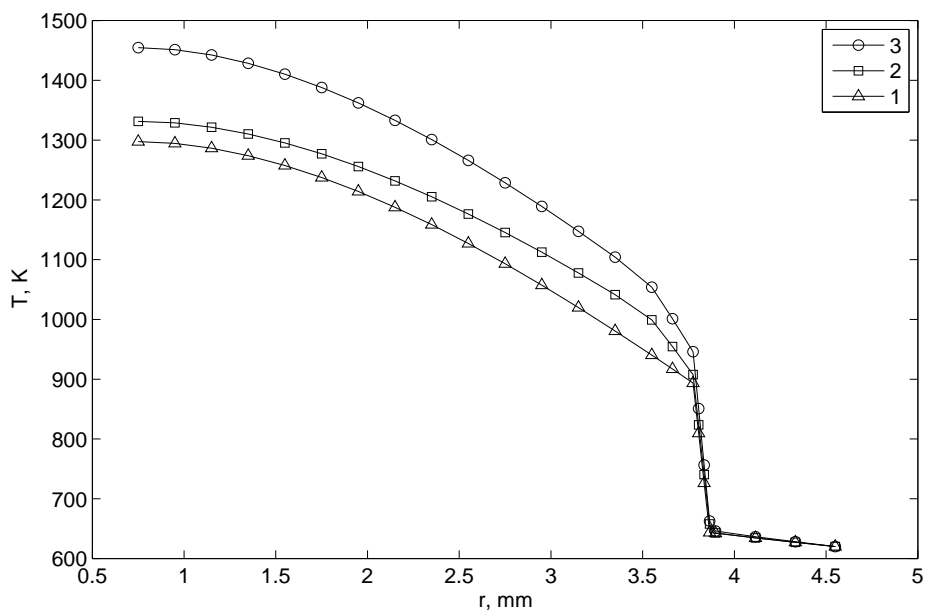,width=140mm}}
\caption{Dependence of temperature on radius at $q_L= 200 $ W/cm for cases:
1:  $bu = 0$; 2:  $bu = 120$ ($q_L=200$ W/cm); 3 :   $bu=120$ ($q_L= 227$ W/cm)}
\label{fig:f6}
\end{figure}

\begin{figure}[ht] 
\centerline{\epsfig{file=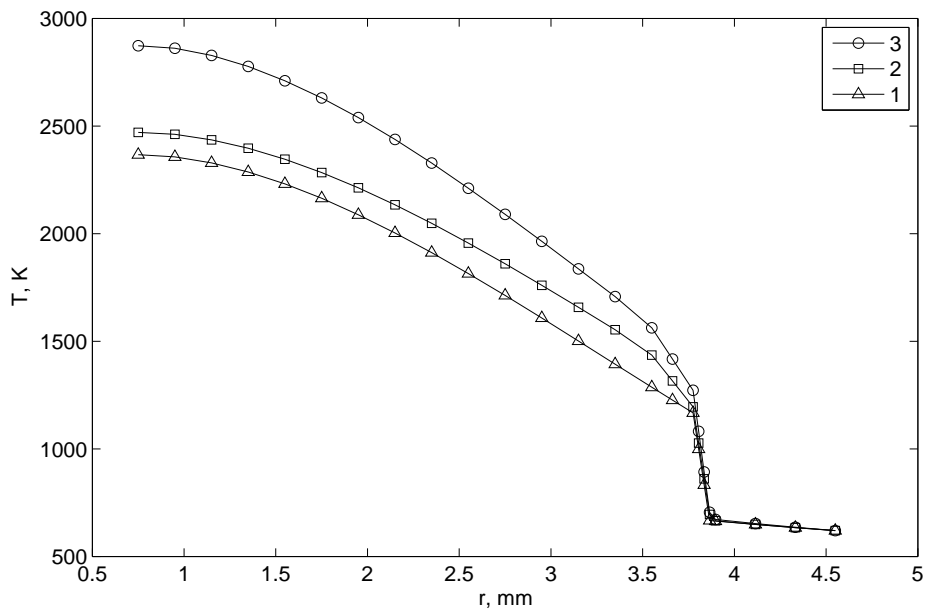,width=140mm}}
\caption{Dependence of temperature on radius at $q=300 $ W/cm$ $ for cases:
1:  $bu = 0$; 2:   $bu = 120$ ($q_L=4$ W/cm); 3:   $bu=120$ ($q_L=453$ W/cm)  }
\label{fig:f7}
\end{figure}

The dependence of temperature on radius in the case of the normal
behavior in nuclear fuel rod and for high burn-up is given in figure
\ref{fig:f6} ($q=0.2$) and in figure \ref {fig:f7} (q=0.4).

We can observe that the high burn-up leads to the increasing of the
temperature in nuclear fuel rod. The point of the melting of $UO_2$ is
equal to 3000 K. In the case of the high burn-up equal to 120 we can
believe that the fuel can melt and we need to decrease the
power of the source in the nuclear fuel rod.

\begin{figure}[ht] 
\centerline{\epsfig{file=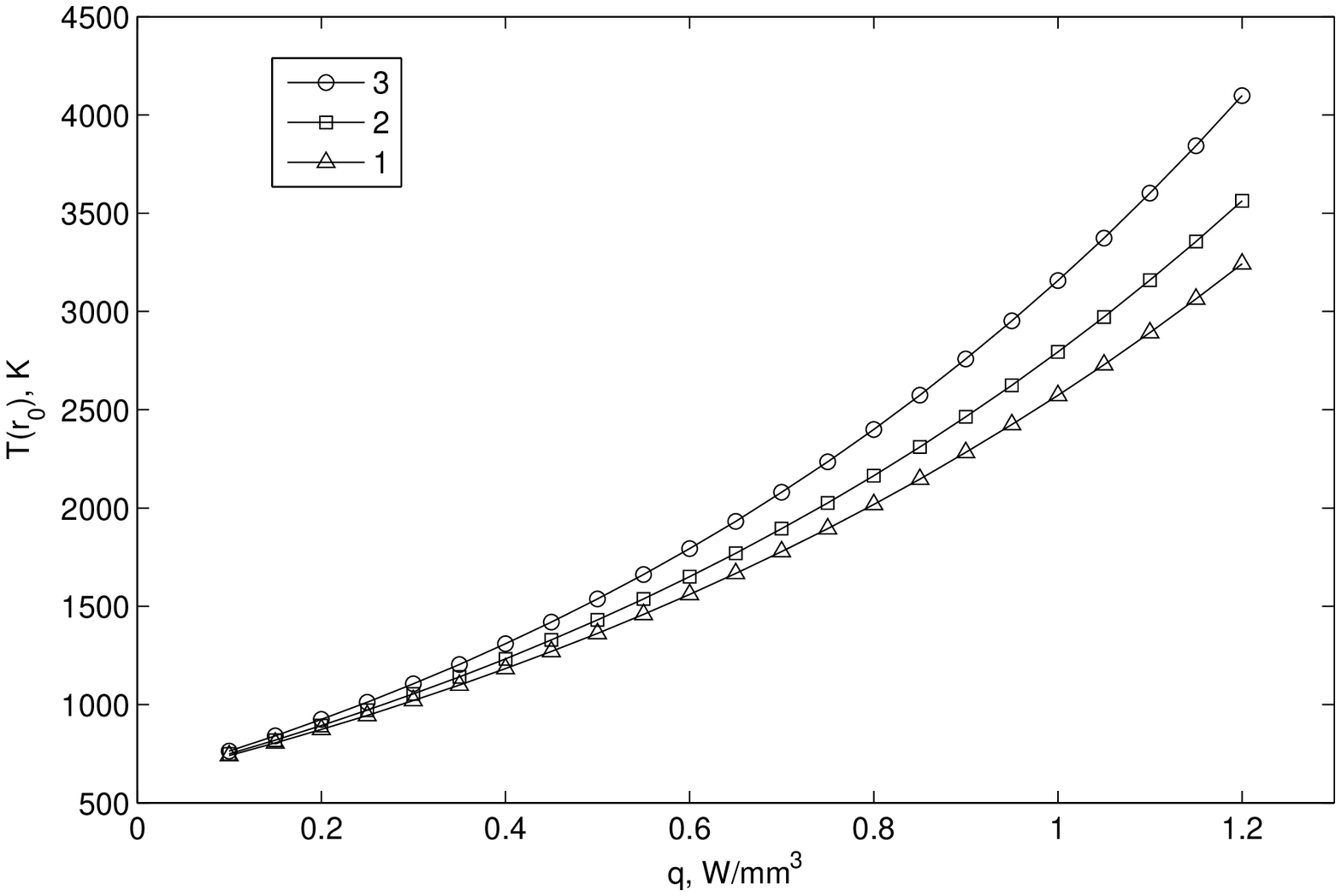,width=140mm}}
\caption{Dependence of temperature in point  $r=r_0$ on linear power of nuclear
fuel rod  ($q_1 = 2 q$) at
different burn-up 1:  $bu=0$; 2:  $bu=60$; 3 :  $bu=120$;}
\label{fig:f8n3}
\end{figure}

Dependence of temperature on source $q$ in the point $r=r_0$ of the
nuclear fuel rod is presented in figure \ref{fig:f8n3} at various
depths of burn-up of the fuel.

From this figure we can see how should we have to increase the power in
nuclear fuel rod to have the normal behavior of the fuel element.
The formation of the rim - layer leads to increasing of the heat
conductivity. As this fact takes place the power of the source
in nuclear fuel rod is decreased. The linear power of the nuclear
fuel rod is decreased as well. Temperature in the fuel and in the
rim - layer is decreased. We can see this process in figure
\ref{fig:f6} and in figure \ref{fig:f7}. Therefore we have to
increase the power in nuclear fuel rod.

\section{Conclusion}

Let us shortly formulate results of this paper. We have studied the
temperature distribution in nuclear fuel rod taking into account the
fuel, the rim - layer, the gap, the film of the zirconium oxide and
the cladding. Solving the nonstationary task of the temperature
distribution in the nuclear fuel rod by the numerical simulation  we
have obtained that the evolution of the temperature in the rod to
the stationary behavior goes the short time. This time is less then
45 seconds for the maximum power in nuclear fuel rod. This fact
allowed us to consider the solution of the stationary behavior of
the nuclear reactor. We  have solved this task using the analytical
method and we have found the exact solution of the temperature
distribution in the nuclear fuel rod. We analyzed the different
thermal regimes for the stationary behavior reactor and have shown
the important role of the rim - layer of high burn-up nuclear fuel
rod.

\end{document}